\documentclass[conference]{IEEEtran}
\IEEEoverridecommandlockouts
\usepackage{cite}
\usepackage{amsmath,amssymb,amsfonts}
\usepackage{algorithmic}
\usepackage{graphicx}
\usepackage{textcomp}
\usepackage{xcolor}
\usepackage{listings}
\usepackage{changepage}
\def\BibTeX{{\rm B\kern-.05em{\sc i\kern-.025em b}\kern-.08em
    T\kern-.1667em\lower.7ex\hbox{E}\kern-.125emX}}

\begin{document}

\title{TS-Verkle: A TypeScript Native Verkle Library With On-chain Verifier \\
}

\author{
\IEEEauthorblockN{Zhikai Li\IEEEauthorrefmark{1}\thanks{}, Xuekai Liu\IEEEauthorrefmark{1}\thanks{*These authors contributed equally to this work.} Boyuan Xu, Eric Chen, and Bhaskar Krishnamachari}
\IEEEauthorblockA{\textit{Viterbi School of Engineering, University of Southern California, Los Angeles, USA}\\
\{leol, xuekailiu, boyuanxu, ericsc, bkrishna\}@usc.edu}
}

\maketitle

\begin{abstract}
Blockchain systems face significant scalability challenges due to growing data volumes and increasing transaction demands, necessitating more efficient data structures and verification mechanisms. Verkle trees, a novel data structure combining the efficiency of Merkle trees with the compactness of vector commitments, have gained attention for their potential to optimize blockchain storage and improve scalability. However, their practical implementation, especially at the smart contract level, has remained unexplored. To address these challenges, we present TS-verkle, the first known TypeScript-native implementation of Verkle trees designed for web3 backend compatibility, coupled with a corresponding on-chain verifier written in Solidity. 
Our work bridges this gap by providing a concrete implementation of Verkle trees and demonstrating their feasibility for on-chain verification. While previous literature suggests Verkle trees should outperform Merkle trees due to their succinct proof size, our empirical evaluation reveals that basic implementations of Verkle trees actually incur higher costs than Merkle trees without advanced optimization techniques. This finding represents a crucial insight for blockchain developers and researchers considering Verkle tree adoption. The paper discusses implementation strategies and performance characteristics while exploring implications for scaling and data availability in decentralized blockchain systems.

\end{abstract}


\vspace{-0.3cm}

\section{Introduction}
Data integrity and validation in distributed systems have become increasingly critical challenges in modern computing. The Merkle tree, proposed by Ralph C. Merkle in 1979, emerged as a fundamental data structure addressing these challenges through its innovative approach to data verification\cite{merkle1987digital}. Built upon the foundation of binary search trees, Merkle trees extend traditional tree properties by incorporating hash values at each node: non-leaf nodes contain the hash of their children, while leaf nodes store the hash of their corresponding data\cite{becker2008merkle}\cite{9545917}. This structure makes it possible to verify whether a leaf is part of the tree without leaking any other data except the hash value. Bitcoin, introduced by Satoshi Nakamoto in 2008, demonstrates an important application of Merkle trees. In the Bitcoin system, transactions are stored in Merkle trees, allowing block headers to contain just the root hash value to verify transactions. This approach eliminates the need to store large amounts of transaction data while maintaining verification capabilities. A detailed explanation of the verification in the Merkle tree can be found in \cite{9545917}\cite{9588047}\cite{berman2007optimal}.



However, despite their efficient proof mechanism using hashes instead of complete information, Merkle trees face a significant limitation: proof size scalability. Merkle tree proofs need nodes at every level, making the proof size dependent on the tree's depth. For a standard binary Merkle tree containing $n$ blocks of information, the proof size is $O(\log_2 n)$. For example, a Merkle tree with a depth of just 30 requires about 1 kilobyte of proof data using SHA-256. This limitation has become particularly challenging with the growth of networks like Ethereum, where verifying on-chain data grows increasingly complex. Therefore, reducing the proof size of the Merkle tree becomes an important way to enhance the scalability of the blockchain system.

Early attempts at optimization, such as creating k-ary Merkle trees to reduce the number of layers, actually increased proof sizes from $O(\log_2 n)$ to $O(k\log_k n)$. Researchers have proposed various solutions: Ayyalasomayajula and Ramkumar proved that SubTree implementation works better than LinearTree for large relational databases\cite{10438756}, while Mizrahi et al. developed a traffic-aware Merkle tree that organizes data based on transaction patterns to reduce average proof length\cite{9352820}.

\begin{figure}[htbp]
\centering
\includegraphics[scale=0.29]{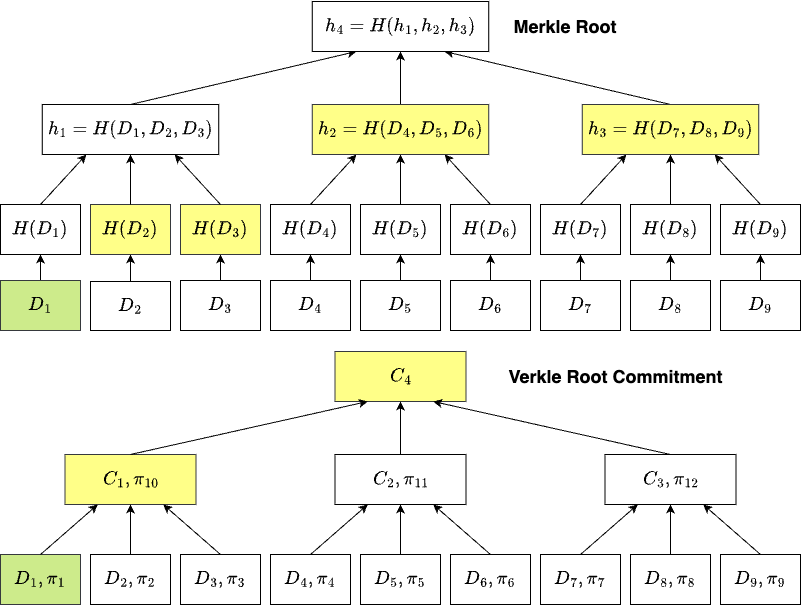}
\caption{A 3-ary Merkle tree (above) and a 3-ary Verkle tree (below), with proofs highlighted in yellow}
\label{fig:Merkle-Verkle}
\end{figure}

Verkle tree emerges as a new data structure that significantly reduces proof sizes. The core idea is to use vector commitments instead of hash functions in $k$-ary Merkle trees, greatly reducing the size of proofs and making trade-offs between efficiency\cite{kuszmaul2019verkle}. Specifically, in Merkle trees, the parent node is the hash value of its child nodes, while in Verkle trees, it is the vector commitment of its child nodes. Thus, the Verkle tree only needs to provide the path and minimal additional information for verification, shown in figure \ref{fig:Merkle-Verkle}, eliminating the need for same-level nodes, which greatly reduces the proof size to $O(\log_k n)$. Currently, Verkle tree has been applied to tasks such as post-quantum digital signatures\cite{iavich2023post}, file verification systems\cite{10226788}, reducing block incentive volatility\cite{zhao2024minimizing}, and protecting distributed system logs\cite{boiko2024distributed}.

Despite these advances, there remains a notable gap in research regarding the feasibility of replacing Merkle trees with Verkle trees in on-chain applications, particularly at the smart contract level. To address this gap, we propose TS-Verkle\footnote{Github link will be released upon acceptance}, a TypeScript library for implementing on-chain verifiers based on Verkle trees. The following will detail the TypeScript implementation of the Verkle tree structure and the smart contract that performs on-chain proof verification. We hope that this library will further promote the application of Verkle trees in on-chain verification and provide a reference to improve the scalability of blockchain systems.

Our contributions can be summarized as follows:
\begin{itemize}
    \item We introduce the first known TypeScript Native Verkle Tree library with on-chain verifier
    \item We provide experimental results on Merkle Tree vs Verkle Tree on-chain verification gas cost
    \item We introduce Ethereum Gas Cost Equations dependent on various Verkle Tree and Merkle Tree configurations 
    \item We empirically demonstrate that Verkle trees incur higher gas costs, contradicting prior theoretical efficiency assumptions.
\end{itemize}


\section{Related Work}
This section examines existing Verkle tree implementations and analyzes the polynomial commitment mechanisms that enable Verkle trees' functionality.

\subsection{Verkle Tree Implementations}
Current Verkle tree implementations span a variety of programming languages, each suitable for different applications, but no project has compared both Merkle trees and Verkle trees. Notably, no language includes an on-chain validator. Our TS-Verkle implements on-chain integration, solves the on-chain verification gap, and provides a side-by-side comparison between Merkle trees and Verkle trees.

\subsubsection{Rust implementation}
 Rust implementation \textit{rust-verkle}\cite{rust-verkle} leverages the language's safety guarantees and performance optimizations. Its low-level memory management capabilities and fast cryptographic libraries optimize polynomial commitments and curve algorithms.

\subsubsection{Golang implementation}
The Ethereum community's \textit{go-verkle} \cite{go-verkle} exemplifies Go's focus on simplicity and concurrency for blockchain nodes. While offering native concurrency support, Go's garbage collection creates minor performance tradeoffs compared to Rust implementations.

\subsubsection{Python implementation}
Python's \textit{verkle-trie-ref}\cite{verkle-trie-ref} serves primarily for prototyping and research. It benefits from extensive library support, including \textit{py\_ecc} for elliptic curve operations and \textit{numpy} for polynomial processing, making it valuable for theoretical validation.

\subsection{Polynomial Commitment(PC) Mechanisms}
Polynomial commitments (PC) are often used as a vector commitment mechanism for Verkle trees. PC schemes enable a prover to commit to a polynomial $f(x)$ and then provide a succinct proof that the polynomial evaluates to the claimed value $y$ at a given point $x=z$. In a Verkle tree, nodes are represented as vector $v=[v_1,v_2,...,v_n]$ and encoded as a polynomial $f(x)$. Commitments are made to vectors using polynomial commitments. Two widely used PC schemes are KZG commitments and Inner Product Arguments (IPA).

\subsubsection{KZG Polynomial Commitments}
The KZG commitment scheme, named after Kate, Zaverucha, and Goldberg, is a popular PC mechanism for Verkle tree\cite{kate2010constant}. It's security relies on elliptic curve pairings. The size of KZG proofs is constant and independent of the polynomial degree. KZG is favored by blockchain systems such as Ethereum proto-danksharding (EIP-4844) due to its efficiency and simplicity. There are already open-source KZG libraries based on TypeScript, such as libkzg\cite{libkzg}.

\subsubsection{Inner Product Argument (IPA) Commitments}
IPA-based commitments use recursive inner products to achieve succinctness\cite{bootle2016efficient}. It does not require a trusted setup, relying instead on discrete logarithm assumptions. So, the proof size is dependent on the polynomial degree. Generating proofs in IPA involves multiple recursive computations, making it computationally expensive compared to KZG. It is often chosen for systems that prioritize transparency and trustlessness, such as zero-knowledge proof protocols.

More comparisons of KZG and IPA are shown in Table \ref{tab:KZG&IPA}. In our TypeScript implementation, we focus on KZG commitments due to their efficiency and simplicity for practical Verkle tree use cases.

\begin{table}[htbp]
\begin{adjustwidth}{0.2cm}{}
\caption{KZG vs. IPA}
\begin{tabular}{|l|c|c|}
\hline
& KZG & IPA\\
\hline
Basic Tech & Elliptic Curve Pairings & Discrete Logarithm \\
Setup Requirements & Trusted Setup Required & No Trusted Setup \\
Transparent & No & Yes \\
Succinct & Yes & No \\
Proof Size  & $O(1)$ & $O(\log n)$ \\
Verify Time  & $O(1)$ & $O(\log n)$ \\
Prove Time & $O(n)$ & $O(n)$ \\
\hline
\end{tabular}
\label{tab:KZG&IPA}
\end{adjustwidth}
\end{table}

\section{Implementation}
The design of TS-Verkle has two major components: the local Verkle tree and the corresponding onchain verifier.

\subsection{Local Verkle Tree}\label{AA}

\begin{figure}[ht]
\centering
\includegraphics[scale=0.7]{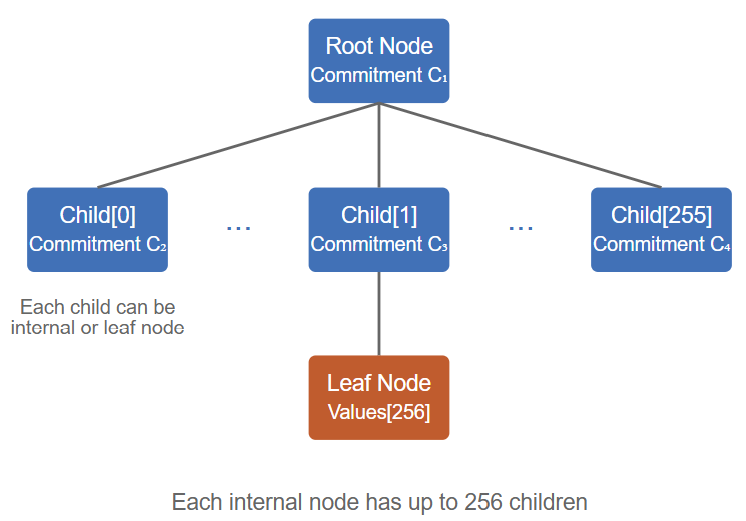}
\caption{Verkle and Merkle Tree gas cost in respect to tree capacity}
\label{fig:tree_structure}
\end{figure}

Our Verkle tree implementation follows a modular architecture with two primary node types: internal nodes and leaf nodes. Each internal node maintains a mapping of up to 256 child nodes, while leaf nodes store the actual values. The implementation leverages the libkzg library \cite{libkzg} for polynomial commitment operations, specifically for generating and verifying single-point proofs.
The core functionality is encapsulated in a \texttt{VerkleTree} class that handles recursive tree traversal, node insertion, and proof generation. Each node implements a common interface that includes methods for commitment generation (\texttt{commit()}), proof generation (\texttt{getProof()}), and value retrieval (\texttt{get()}).
For proof generation, the implementation traverses the path from root to leaf, collecting necessary commitments and generating proofs at each level using KZG polynomial commitments. A Verkle proof consists of several components:

\begin{lstlisting}[basicstyle=\footnotesize ]
interface VerkleProof {
    stem: Uint8Array;        
    leafIndex: number;        
    indices: number[];        
    internalProofs: Proof[];  
    internalChildCommitments: Commitment[];  
    leafProof: Proof;        
}
\end{lstlisting}

The proof structure enables efficient verification by providing all necessary components: the stem-index structure for path representation, where each key is split into a stem portion (determining the path through internal nodes) and a leaf index (identifying the specific value within a leaf node). The verification process recursively validates these proofs by checking the KZG commitments at each level, ensuring the integrity of the path from the root commitment to the claimed value at the leaf node.
\subsection{On-chain Verifier}
The on-chain verifier implements the verification logic in Solidity, following the same mathematical principles as the local verification process. It takes a Verkle proof as input and validates the path from root to leaf by verifying the KZG polynomial commitments at each level. The implementation leverages precompiled contracts for efficient elliptic curve operations and performs the same recursive validation of internal node proofs and leaf proofs as the local verifier. This ensures that any proof generated by the local Verkle tree can be efficiently verified within the constraints of the Ethereum Virtual Machine.

\section{Experiments}
The experimental framework was structured in two phases: local tree construction and on-chain verification. Our methodology aimed to compare the performance characteristics of Verkle trees against traditional Merkle Trees, specifically utilizing the widely-audited OpenZeppelin implementation as our baseline.

\subsection{Local Tree Construction}
Our experimental setup focused on comparing the efficiency of Verkle trees against OpenZeppelin's Merkle Tree implementation in managing sets of Ethereum addresses. We constructed trees of varying sizes, ranging from small sets of $2^3$ (8) addresses to larger sets of  $2^{20}$ (1048576) addresses. The local construction phase involved building both types of trees and generating membership proofs for random addresses within the set. The Merkle Tree implementation followed OpenZeppelin's standard binary tree structure. Since Verkle tree proof size is independent of the branching factor, our Verkle tree implementation utilizes a branching factor of 256 because it is the most commonly used value.

\subsection{On-chain Verification}
The on-chain verification phase concentrated on measuring the gas costs associated with verifying individual element proofs. We deployed both Verkle and Merkle verification contracts to a local Hardhat network and conducted systematic gas consumption analysis for proof verification operations. The verification process involves confirming that a given address exists within the original set by validating its proof against the tree's root hash. We specifically focused on single-element verification scenarios, as our current implementation does not support batch operations.

\section{Results}


\subsection{Verkle Tree Gas Cost Analysis}

\begin{table}[htbp]
\begin{adjustwidth}{0cm}{}
\caption{Verkle Tree}
\begin{tabular}{|l|c|c|c|c|c|}
\hline
& 2 levels & 3 levels & 4 levels & 5 levels & 6 levels\\
\hline
Capacity  & $256^2$ & $256^3$ & $256^4$ & $256^5$ & $256^6$ \\
Proof length  & constant & constant & constant & constant & constant \\
Verification Cost & 285507 & 428707 & 571144 & 714178 & 864628 \\
Calldata Cost & 25210 & 27770 & 30330 & 32890 & 35450  \\
Total & 491527 & 637287 & 782284 & 927878 & 1080888  \\
\hline
\end{tabular}
\label{tab:VerkleGas}
\end{adjustwidth}
\end{table}

When using Verkle tree for verification, most of the cost comes from verification itself with the remainder going to Calldata. Analysis of the empirical results demonstrates that Verkle tree verification costs follow a logarithmic relationship with respect to capacity. The experimental data presented in Table \ref{tab:VerkleGas} illustrates that while the capacity grows exponentially with each additional level, the verification costs exhibit a linear growth rate, and so does the calldata cost. The total gas cost of verifying a single element from a Verkle tree with $k$ branching factor and $C$ capacity can be modeled by the equation:

\begin{equation}
\left\lceil\log_{k}(C)\right\rceil \times 147560 + 200900\label{verkleCost}
\end{equation}

\subsection{Merkle Tree Gas Cost Analysis}

\begin{table}[htbp]
\vspace{-0.3cm} 
\caption{Merkle Tree}
\begin{adjustwidth}{-0.5cm}{}
\begin{tabular}{|l|c|c|c|c|c|}
\hline
& 3 levels & 7 levels & 10 levels & 15 levels & 20 levels\\
\hline
Capacity & $2^{3}$ & $2^{7}$ & $2^{10}$ & $2^{15}$ & $2^{20}$\\
Proof length  & 3 & 7 & 10 & 15 & 20 \\
Verification Cost & 25610 & 28715 & 31407 & 35602 & 39770 \\
Calldata Cost & 2816 & 4736 & 6386 & 8948 & 11500  \\
Total & 28426 & 33451 & 37793 & 44550 & 51270  \\
\hline
\end{tabular}
\end{adjustwidth}
\label{tab:MerkleGas}
\end{table}

Analysis of the empirical results demonstrates that Merkle tree verification costs follow a logarithmic relationship with base 2, as opposed to base $k$ in Verkle trees. The experimental data presented in Table \ref{tab:MerkleGas} shows that capacity grows exponentially with levels, while both verification costs and proof length increase linearly with tree depth. The verification cost can be modeled by equation:

\begin{equation}
(\left\lceil\log_{2}(C)\right\rceil \times 1342) + 24300\label{merkleCost}
\end{equation}

\subsection{Comparative Analysis}
\vspace{-0.3cm} 
\begin{figure}[ht]
\centering
\includegraphics[scale=0.4]{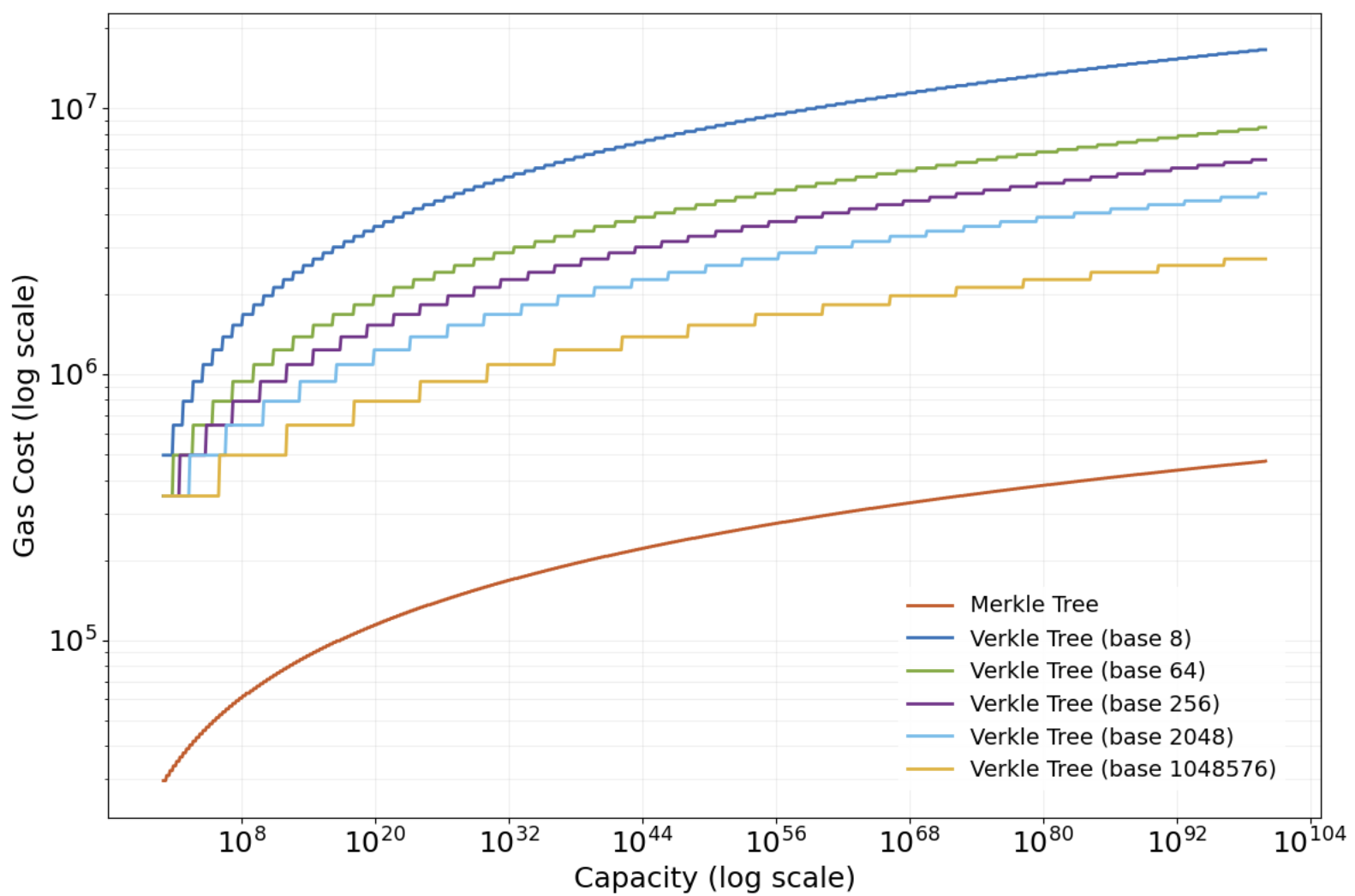}
\caption{Verkle and Merkle Tree gas cost in respect to tree capacity}
\label{fig:compare_cost}
\end{figure}

    Although~\cite{kuszmaul2019verkle} has anticipated that Verkle trees could be more efficient for larger datasets due to their constant-size proofs, figure \ref{fig:compare_cost}, produced using \eqref{merkleCost} and \eqref{verkleCost}, reveals a different outcome in terms of on-chain gas costs. The data shows that Merkle Trees consistently maintain lower total gas costs across all capacity ranges tested, contradicting initial theoretical expectations.

This unexpected result can be attributed to several factors:

\begin{enumerate}
    \item The significantly higher base verification cost for Verkle trees (285,507 gas at 2 levels) compared to Merkle Trees (25,610 gas at 3 levels) creates a substantial initial overhead that their constant-size proof advantage doesn't overcome.
    
    \item While Verkle trees achieve better capacity scaling ($k^n$ vs $2^n$) and constant proof sizes, the actual verification computation on-chain proves to be more expensive than handling larger proof sizes in Merkle Trees.
    
    \item The linear growth in Merkle Tree costs (approximately 1,342 gas per level) proves to be more economical than Verkle tree's higher per-level cost (around 147,560 gas per level), even when accounting for increased calldata costs from larger proof sizes.
\end{enumerate}

This analysis suggests that while Verkle trees offer theoretical advantages in terms of proof size and capacity scaling, their current implementation results in higher on-chain computational costs that outweigh these benefits. For blockchain applications where gas optimization is crucial, Merkle Trees remain the more cost-effective solution across all tested capacity ranges.


\section{Conclusion}
The emergence of the Verkle tree as a new data structure marks major progress in optimizing blockchain system structure. It uses vector commitment instead of hash function to greatly reduce the proof size while considering the proof efficiency. To fill the application gap of Verkle tree in on-chain verification, this paper proposes TS-Verkle, a Verkle tree library developed based on TypeScript and including on-chain validator. Through experiments, we compared the differences between Verkle tree and Merkle Tree in on-chain verification gas cost. Our development of Ethereum Gas Cost Equations reveals an important insight: while Verkle trees demonstrate theoretical advantages, their current implementation faces challenges in achieving superior gas cost efficiency compared to Merkle trees.

The path forward for realizing the full potential of Verkle trees requires advancement along multiple research directions. First, exploring alternative vector commitment schemes beyond KZG and IPA presents promising opportunities. Transparent commitments and grid-based commitments could potentially reduce verification times while eliminating the need for trusted setup procedures. Second, optimization of on-chain verification demands investigation into advanced techniques such as batch verification and the implementation of precompiled contracts for elliptic curve operations, both of which could significantly reduce gas costs. Back to the tree structure, potential solutions include hybrid implementations that combine the strengths of both Merkle and Verkle trees, as well as adaptive methodologies capable of dynamic tree structure optimization \cite{kuznetsov2024adaptive}. Looking ahead, Verkle trees hold considerable promise in improving blockchain scalability. Through our research and implementation of TS-Verkle, we have established a foundation for future exploration and development in this field. We hope that our work paves the way for continued innovation in the efficiency and scalability of decentralized systems.
 
\section*{Acknowledgement}
This work was supported in part by a grant from AFOSR. 


\newpage
\bibliographystyle{IEEEtran}
\bibliography{bibliography}

\end{document}